\documentclass[aps,secnumarabic,twocolumn,nobalancelastpage,amsmath,amssymb,nofootinbib,floatfix]{revtex4}
\usepackage{amsfonts}
\usepackage{amsmath}
\usepackage{amssymb}
\usepackage[colorlinks, citecolor=blue,linkcolor=blue
]{hyperref}
\usepackage{color}
 \usepackage{float}
\usepackage{textcomp}
\usepackage{subfigure}
\usepackage{graphicx}
\begin{document}
 \title{Steady-state entanglement of Bose-Einstein condensate and a nanomechanical resonator} 
\author{Muhammad Asjad}
\email{asjad\_qau@yahoo.com}
\author{Farhan Saif}
\email{fsaif@yahoo.com}
\affiliation{Department of Electronics, Quaid-i-Azam University, 45320 \ Islamabad, Pakistan.}
\begin{abstract}
We analyze the steady-state entanglement between Bose-Einstein condensate trapped inside an optical cavity with a moving end mirror (nanomechanical resonator) driven by a single mode laser. The quantized laser field mediates the interaction between the Bose-Einstein condensate and nanomechanical resonator. In particular, we study the influence of temperature on the entanglement of the coupled system, and note that the steady-state entanglement is fragile with respect to temperature.        
\end{abstract} \maketitle
\section{\bf{Introduction}}
 Optical nano-mechanical systems that couple optical degree of freedom to the mechanical motion of a cantilever have been subject of increasing investigation \cite{Kipp}. In these optomechanical systems coupling is obtained via radiation pressure inside a cavity \cite{Braginsky, Mancini, Zhang}, or via quantum dots \cite{Tian} or ions \cite{Naik}. Recently, it is made possible to couple mechanical resonators with the ensembles of atoms, where, the interaction is mediated by the field inside the cavity which couples the mechanical resonators to the internal degrees of freedom of the atoms \cite{Ian,genes}, or to motional degrees of the freedom of the atoms \cite{Meiser} causing effects, (\textit{e.g}, cooling of the mechanical resonator via bath of atoms \cite{Ritsch}). In quantum meteorology, various targets, such as, measurement of displacement with larger accuracy \cite{Rugar} and the detection of gravitational waves \cite{V. Braginsky} are set as milestone achievements. Engineering entanglement in nano-mechanical systems is useful in quantum technologies \cite{Nielsen}. The possibility of entangling the electromagnetic field with motional degree of freedom of mechanical systems have been explored in various approaches \cite{Vitali,Paternostro}.  In this paper, we consider a hybrid optomechanical system which consists of a Bose-Einstein condensate (BEC) trapped inside a Fabry-Perot cavity with a vibrating end mirror (nano-mechanical resonators) driven by a single mode optical field. The intracavity field mediates the interaction of nanomechanical resonator with collective oscillations of the atomic density. Hence, the motional degrees of the nano-mechanical resonator indirectly couples to motional degrees of the freedom of the Bose-Einstein condensates via optical field inside the cavity. Therefore, intracavity field acts as nonlinear spring between collective atomic density and nano-mechanical resonator. We show that (i) the mechanical vibration of the nanomechanical resonator is entangled to the motional degree of the freedom of the BEC in the steady state. (ii) Furthermore, we analyze the steady-state entanglement as a function of temperature, coupling strength between BEC and field, moving mirror and field, and power of input driving laser field. In section 2, we model the system and explain its interaction with the environment. In section 3, we calculate the quantum Langevin equations for our system and solve the dynamics. In section 3, we quantify the steady-state entanglement between mechanical resonator and Bose-Einstein condensate. Later, in section 5, we explain the possibility to experimentally measure the generating entanglement. Finally, in section 6, we provide concluding remarks.    
\section{\bf{The Model}}
 We consider a Fabry-Perot cavity with a moving end mirror driven by a single-mode optical field of frequency $\omega_{\mathrm{P}}$, and BEC of $N$-two level atoms are trapped inside the Fabry-Perot cavity \cite{Esteve, Ritter}. The condensate atoms placed in the cavity observe one-dimensional optical lattice,  formed by the oppositely propagated electromagnetic field inside the cavity. We consider that the atom-field detuning $\Delta_{\mathrm{a}}$ is very large, spontaneous emission is negligible, and, as a consequence we adiabatically eliminate the internal excited state dynamics of the atoms. In addition, we also consider that the atomic densities are low enough that one can neglect the two-body interactions. In a weakly interacting regime, the recent experiment \cite{Ritter} suggests that only the first two symmetric momentum side modes are excited with momentum $\pm \,2 \mathrm{\hbar}$k, where k is the wave number of the field. Moreover, we assume that the vibrating end mirror of the optical cavity of length $L$ is performing harmonic oscillations with frequency $\omega_{\mathrm{m}}$ along the cavity axis (x-axis).
 \paragraph*{} The Hamiltonian of the hybrid optomechanical system formed by the BEC, the intracavity field, and the vibrating end mirror of the cavity (nanomechanical resonator) in the rotating frame at the laser frequency $\omega_{\mathrm{p}}$ is given by \cite{Nagy,cklaw}
 \begin{equation}
\hat{\mathrm{H}}=\sum_{\mathrm{i=\,a,\,b\,c}} \hat{\mathrm{H}}_{\mathrm{i}} +\hat{\mathrm H}_{\mathrm{ac}}+\hat{\mathrm H}_{\mathrm{mc}}\,, \label{ham} 
  \end{equation}
where, $\hat{\mathrm{H}}_{\mathrm{a}}=\mathrm{\hbar}\, \omega_{\mathrm a}\, \hat{\mathrm{a}}^\dag\hat{\mathrm{a}}$, $\hat{\mathrm{H}}_{\mathrm{b}}=\mathrm{\hbar}\, \omega_{\mathrm{m}}\, \hat{\mathrm{b}}^\dag\hat{\mathrm{b}}$ and  $\hat{\mathrm{H}}_{\mathrm{c}}=\mathrm{\hbar}\, \Delta\, \hat{\mathrm{c}}^\dag\hat{\mathrm{c}}-i\,\mathrm{\hbar}\,\mathrm{E}\,(\hat{\mathrm c}-\hat{\mathrm c}^\dag)$. Here $\hat{\mathrm c}\,(\hat{\mathrm c}^\dag$) is the annihilation (creation) operator of the single-mode optical field, $\Delta= \omega_{\mathrm c}-\omega_{\mathrm p}+\mathrm{N} \mathrm{U}_{\mathrm{o}}/2$, where $\mathrm{U}_{\mathrm{o}}$ is the optical lattice barrier depth per photon and represents the atomic back action on the field \cite{Maschler}, and $-i\,\mathrm{\hbar}\,\mathrm{E}\,(\hat{\mathrm c}-\hat{\mathrm c}^\dag)$ shows the driving of the cavity field with amplitude $\mathrm{E}$ related to the laser power $\mathrm{P}$ by $|\mathrm E|=\sqrt{2\,\mathrm{P}\, \kappa/\mathrm{\hbar}\, \omega_{\mathrm{p}}}$, where $\kappa$ is the decay rate of the photons in to the associated outgoing modes. Moreover, $\hat{\mathrm{a}}\, (\hat{\mathrm{a}}^\dag)$ is the annihilation (creation) operator of the BEC mode with frequency $\omega_{\mathrm a}=4 \omega_{\mathrm r}$, where $\omega_{\mathrm r}$ is the recoiled frequency. In addition, $\omega_{\mathrm m}$ and $\hat{\mathrm{b}}\,(\hat{\mathrm{b}}^\dag)$ are frequency and annihilation (creation) operator of the nano-mechanical resonator, respectively. We assume that the frequency of the nanomechanical resonator is less than the free spectral range, (\textit{i.e}, $\omega_{\mathrm{m}}<<\mathrm{v}/2\,\mathrm L\,$ (v is the speed of light). Therefore, scattering of photons into others modes, except the driven mode, is neglected \cite{cklaw}. Here,  $\hat{\mathrm H}_{\mathrm{mc}}$ accounts for the interaction between nano-mechanical resonator and intracavity field and is given by
\begin{equation}
\hat{\mathrm{H}}_{\mathrm{mc}}=-i\mathrm{\hbar}\, \dfrac{\mathrm{g}_{\mathrm{mc}}}{\sqrt{2}}\,\hat{\mathrm{c}}^\dag\hat{\mathrm{c}} \,(\hat{\mathrm b}+\hat{\mathrm b}^\dag)\,. \label{hmc}
\end{equation} 
Here, the coupling strength between nano-mechanical resonator and light radiation pressure is defined by $\mathrm{g}_{\mathrm{mc}}=\sqrt{2}(\omega_{\mathrm p}/\mathrm L)\,\mathrm{x}_{\mathrm o}$, where $\mathrm{x}_{\mathrm o}=\sqrt{\mathrm{\hbar}/2\,\mathrm{m}\,\omega_{\mathrm{m}}}$ is the zero-point motion of the mechanical mode of mass $\mathrm{m}$. Moreover, the interaction of the BEC with intracavity mode is described by the  Hamiltonian $\hat{\mathrm H}_{\mathrm{ac}}$ given by
  \begin{equation}
\hat{\mathrm{H}}_{\mathrm{ac}}=i\mathrm{\hbar}\, \dfrac{\mathrm{g}_{\mathrm{ac}}}{\sqrt{2}}\,\hat{\mathrm{c}}^\dag\hat{\mathrm{c}} \,(\hat{\mathrm a}+\hat{\mathrm a}^\dag)\,, \label{hac}
\end{equation}
where, $\mathrm{g}_{\mathrm{ac}}=(\mathrm{U}_{\mathrm{o}}\sqrt{\mathrm {N}})/2$ is described the strength of interaction between BEC mode and intracavity field. 
\paragraph*{} In order to describe the complete dynamics of the system we include the dissipation effects. In addition to the dynamics described by the Hamiltonian in Eq.(\ref{ham}), the system is exposed to the random noise forces due to quantum fluctuations of the radiation field and fluctuation of the phononic heat bath associated to the mechanical resonator. We neglect the thermal effects of the atomic cloud and assume that the vacuum noise associated with the cavity field is Markovian in nature with decay rate $\kappa$ and noise operator $\hat{\mathrm{c}}_{\mathrm{in}}(t)$ of the input field which obeys the following correlation functions,
\begin{eqnarray}
\left < \hat{\mathrm{c}}_{\mathrm{in}}^\dag(t)\,\hat{\mathrm{c}}_{\mathrm{in}}(t')\right >&=& \mathrm{n}_{\mathrm{c}}\, \delta\left(t-t'\right)\,,\nonumber\\ 
\left < \hat{\mathrm{c}}_{\mathrm{in}}(t)\,\hat{\mathrm{c}}_{\mathrm{in}}^\dag(t')\right >&=& \left(\mathrm{n}_{\mathrm{c}}+1\right)\, \delta\left(t-t'\right). \label{fn}
\end{eqnarray}  
Here, $\mathrm{n}_{\mathrm{c}}=\left[\mathrm{exp}\{\mathrm\hbar \,\omega_{\mathrm c}/K_{\mathrm{B}}\,\mathrm{T}\}-1 \right]^{-1}$ is the equilibrium occupation number of the optical oscillator. For optical frequency $\omega_{\mathrm{c}}$ we consider $\mathrm\hbar \,\omega_{\mathrm c}/K_{\mathrm{B}}\,\mathrm{T}>>1$ and set $\mathrm{n}_{\mathrm c}=0$.
\paragraph*{} The motion of the nano-mechanical resonator is affected due to thermal bath is Brownian and non-Morkovian in nature \cite{genes}. The quantum effects on mechanical resonator are only observed in the limit of very high mechanical quality factor $\mathrm Q=\omega_{\mathrm{m}}/\gamma>>1$, the Brownian noise operator can be modeled as Markovian with the decay rate of the mechanical mode is $\gamma$. Therefore, the noise operator $\hat{\mathrm{b}}_{\mathrm{in}}(t)$ can be characterized as
\begin{equation}
\left < \hat{\mathrm{b}}^\dag_{\mathrm{in}}(t)\,\hat{\mathrm{b}}_{\mathrm{in}}(t')\right >= \mathrm{n}_{\mathrm{th}}\, \delta\left(t-t'\right)\,, \label{mn}
\end{equation}
where, $\mathrm{n}_{\mathrm {th}}= \left[\mathrm{exp}\{\mathrm\hbar \,\omega_{\mathrm m}/K_{\mathrm{B}}\,\mathrm{T}\}-1 \right]^{-1}$ is the equilibrium thermal occupation number of the mechanical resonator.                                                      
\section{\bf{Heisenberg-Langevin equations}}In order to describe the complete dynamics of the subsystems involved in this problem, an adequate choice is to use the formalism of the quantum Langevin equations. Therefore, the Heisenberg-Langevin equation of motion for the intracavity mode, mechanical mode and bosonic field operator can be written as
\begin{eqnarray}
\dot{\mathrm{a}}&=& -i\,\omega_{\mathrm{a}}\,\mathrm{a}-i\,\dfrac{\mathrm{g}_{\mathrm{ac}}}{\sqrt{2}}\,\mathrm{c}^\dag \mathrm{c} \,, \nonumber \\
\dot{\mathrm{b}}&=& -i\omega_{\mathrm{m}}\,\mathrm{b}+i\,\dfrac{\mathrm{g}_{\mathrm{mc}}}{\sqrt{2}}\,\mathrm{c}^\dag \mathrm{c} -\gamma\,\mathrm{b}+\,\sqrt{2\,\gamma}\,\mathrm{b}_{\mathrm{in}}\,,\nonumber  \\ 
\dot{\mathrm{c}}&=& \left(-i\Delta_{\mathrm{o}} +i\,\dfrac{\mathrm{g} _{\mathrm{mc}}}{\sqrt{2}}\,(\mathrm{b}+\mathrm{b}^\dag)-i\,\dfrac{\mathrm{g} _{\mathrm{ac}}}{\sqrt{2}}\,(\mathrm{a}+\mathrm{a}^\dag)-\kappa\right)\,\mathrm{c} \nonumber \\
&& + E + \sqrt{2\kappa}\,\mathrm{c}_{\mathrm{in}}\,,\label{langeq} 
\end{eqnarray}
where \textit{dot} denotes the time derivatives and for simplicity we omit the \textit{hat} symbol from the operators. These are the nonlinear quantum Langevin equations and dynamics is complicated. In the following we linearized the operators around the steady state values, $\mathrm{a}=\left< \mathrm{a}\right>_{\mathrm{ss}}+\partial \mathrm{a}$, $\mathrm{b}=\left< \mathrm{b}\right>_{\mathrm{ss}}+\partial \mathrm{b}$, $\mathrm{c}=\left< \mathrm{c}\right>_{\mathrm{ss}}+\partial \mathrm{c}$. Here, we assume that the fluctuation operators $\partial \mathrm{a}$, $\partial \mathrm{b}$ and $\partial \mathrm{c}$ have zero mean. The steady state value of the intracavity mode is $\left<\mathrm{c}\right>_{\mathrm{ss}}=\mathrm{E}/(\kappa+i\,\Delta)$, where the total effective detuning is 
\begin{equation}
\Delta=\Delta_{\mathrm o}-\dfrac{\,\omega_{\mathrm{m}}\,\mathrm{g}^2_{\mathrm{mc}}}{\gamma^2+\omega^2_{\mathrm{m}}}\left<\mathrm{c^\dag c}\,\right>_{\mathrm{ss}}-\dfrac{\,\mathrm{g}^2_{\mathrm{ac}}}{\omega_{\mathrm{a}}}\,\left< \mathrm{c^\dag c}\right>_{\mathrm{ss}}. \label{detuning}
\end{equation} 
For the sake of simplicity we assume that the field is real positive and this can be achieved by adjusting the phase of the laser field. Similarly, the steady state value of the BEC and mechanical resonator modes are $\left<\mathrm{a}\,\right>_{\mathrm{ss}}=[-\mathrm{g}_{\mathrm{ac}}/\sqrt{2}\,\omega_{\mathrm{a}}]\left<\mathrm{c^\dag c}\,\right>_{\mathrm{ss}}$ and $\left< \mathrm{b}\right>_{\mathrm{ss}}=\left[i\,\mathrm{g}_{\mathrm{mc}}/\sqrt{2}(\gamma+i\,\omega_{\mathrm m})\right]\left<\mathrm{c^\dag c}\,\right>_{\mathrm{ss}}$ respectively.
\paragraph*{} We linearize the Langevin equations of motion given in Eq.(\ref{langeq}), and assume that pump field is intense and keep terms only up to first order in the fluctuation operators. We rewrite each Heisenberg operator in Eq.(\ref{langeq}) as a sum of steady state value and fluctuation operator with zero mean value. Therefore, the linear set of equations are,
\begin{eqnarray}
 \partial\dot{\mathrm{a}} &=&  -i\omega_{\mathrm{a}}\partial\mathrm{a}-i\dfrac{\mathrm{G}_{\mathrm{ac}}}{2}\left(\partial\mathrm{c}+\partial\mathrm{c}^\dag\right)\,,\nonumber \\
 \partial\dot{\mathrm{b}} &=& -(\gamma+i\omega_{\mathrm{m}})\partial\mathrm{b} + i\dfrac{\mathrm{G}_{\mathrm{mc}}}{2}\left(\partial\mathrm{c}+\partial\mathrm{c}^\dag\right)+\sqrt{2\gamma}\,\mathrm{b}_{\mathrm{in}}\,,\nonumber  \\
\partial\dot{\mathrm{c}} &=& -\left(\kappa+i\,\Delta \right)\partial \mathrm{c} +i\dfrac{\mathrm{G}_{\mathrm{mc}}}{2}\left(\partial\mathrm{b}+\partial\mathrm{b}^\dag\right)\nonumber\\
&&-i\dfrac{\mathrm{G}_{\mathrm{ac}}}{2}\left(\partial\mathrm{a}+\partial\mathrm{a}^\dag\right)+\sqrt{2\,\kappa}\, \mathrm{c}_{\mathrm{in}}\,.\label{linrz}
\end{eqnarray}
 The linearized quantum Langevin equations show that the fluctuations of mechanical resonator and BEC are now coupled to the cavity field quadrature fluctuations by the effective couplings $\mathrm{G}_{\mathrm{mc}}=\sqrt{2}\,\mathrm{g}_{\mathrm{mc}}\left<\mathrm c\right>_\mathrm{ss}$ and $\mathrm{G}_{\mathrm{ac}}=\sqrt{2}\,\mathrm{g}_{\mathrm{ac}}\left<\mathrm c\right>_\mathrm{ss}$, which can be made very large by increasing the amplitude $\left<\mathrm c\right>_\mathrm{ss}$ of the intracavity field. Linearized quantum Langevin equations (\ref{linrz}) and their corresponding Hermitian conjugate form a system of six first-order coupled operator equations, for which the Ruth-Hurwitz criteria implies that the system will be stable only if the following stability condition is satisfied
\begin{equation}
 \left(\Delta^2+\kappa^2\right)\omega_{\mathrm{m}}\omega_{\mathrm{a}}-\Delta\left(\mathrm{G}^2_{\mathrm{mc}}+\mathrm{G}^2_{\mathrm{ac}}\right)>0\,.\label{stblty}
\end{equation}
In the following we consider only red-detuning regime ($\Delta>0$) and from now on we assume that the above stability condition is satisfied. The quantum Langevin equations (\ref{linrz}) are linear in creation and annihilation operators. We transform to the quadratures, \textit{i.e}, $\partial\mathrm{q}_{\mathrm{m}}=(\partial\mathrm{b}+\partial\mathrm{b}^\dag)/\sqrt{2},\,\partial\mathrm{p}_{\mathrm{m}}=(\partial\mathrm{b}-\partial\mathrm{b}^\dag)/i\sqrt{2},\,\partial\mathrm{q}_{\mathrm{a}}=(\partial\mathrm{a}+\partial\mathrm{a}^\dag)/\sqrt{2},\,\partial\mathrm{p}_{\mathrm{a}}=(\partial\mathrm{a}-\partial\mathrm{a}^\dag)/i\sqrt{2},\,\partial\mathrm{q}_{\mathrm{c}}=(\partial\mathrm{c}+\partial\mathrm{c}^\dag)/\sqrt{2},\,\partial\mathrm{p}_{\mathrm{c}}=(\partial\mathrm{c}-\partial\mathrm{c}^\dag)/i\sqrt{2}$. The system of linearized equations of motion can be written in compact matrix form as $\dot {\mathrm{R}}(t)=\mathrm{M}\,\mathrm{R}(t)+\mathrm{F}(t)$, where, $\mathrm{R}=(\partial\mathrm{q}_{\mathrm{m}},\, \partial\mathrm{p}_{\mathrm{m}},\,\partial\mathrm{q}_{\mathrm{a}},\,\partial\mathrm{p}_{\mathrm{a}},\partial\mathrm{q}_{\mathrm{c}},\,\partial\mathrm{p}_{\mathrm{c}})^\mathrm{t}$ is the vector of the quadrature fluctuations, and in superscript $\mathrm{t}$ describes transpose of matrix. Furthermore, $\mathrm{F}$ is the vector corresponding to noises, whereas $\mathrm{M}$ is drift matrix. Since the quantum noises are white in nature and the dynamics is linearized, hence the state of the system will be a zero mean Gaussian state, and therefore completely determined by the covariance matrix $\mathrm{V}_{\mathrm{ij}}=\left<(\mathrm{R}_\mathrm{i}\,\mathrm{R}_\mathrm{j}+\mathrm{R}_\mathrm{j}\,\mathrm{R}_\mathrm{i})-2\mathrm{R}_\mathrm{i}\,\mathrm{R}_\mathrm{j}\right>$. In order to find the steady state covariance matrix $\mathrm{V}$, we solve the linearized quantum Langevin equations as did in \cite{Vitali}. In steady state the covariance matrix fulfills the Lyapunove equation
\begin{equation}
\mathrm{M}\,\mathrm{V}+\mathrm{V}\,\mathrm{M}^{\mathrm{t}}= -\mathrm{V}_{\mathrm{F}}. \label{lymp}
\end{equation}
Eq.(\ref{lymp}) is the linear matrix equation and can straight forwardly be solved. However, the general exact expression is too cumbersome and not reported here. One can extract all the information about the steady state of the system from the correlation matrix.
\section{\bf{Steady state entanglement}} We compute the entanglement between mechanical mode and the BEC mode in steady-state by tracing out the cavity mode. The steady state entanglement is determined by computing the logarithmic negativity $\mathrm{E}_{\mathrm{N}}$ from the corresponding covariance matrix $\mathrm{V}$. We consider an example where the length of the cavity $\mathrm{L}=1\,\mathrm{mm}$ and laser input power $\mathrm{P}= 50 \,\mathrm{mW}$ \cite{Vitali}. The effective mass and resonance frequency of the mechanical resonator are $\mathrm{m}=4\,\mathrm{ng}$ and $\omega_{\mathrm{m}}= 1\, \mathrm{MHz}$. The decay rate of the mechanical resonator is $\gamma= 2\pi\times 100\, \mathrm{Hz}$ and cavity finesse $\mathrm{F}=1\times10^4$. For these values of the parameters, the coupling $\mathrm{g}_{\mathrm{mc}}$ between mechanical resonator is in the order of $\mathrm{kHz}$. Moreover, the interaction of the optical field with Bose-Einstein condensate  is kept small, so that Bogoliubove mode expansion becomes possible. 
 \begin{figure}[t]
  \centering
 \subfigure[]{\includegraphics[scale=.158]{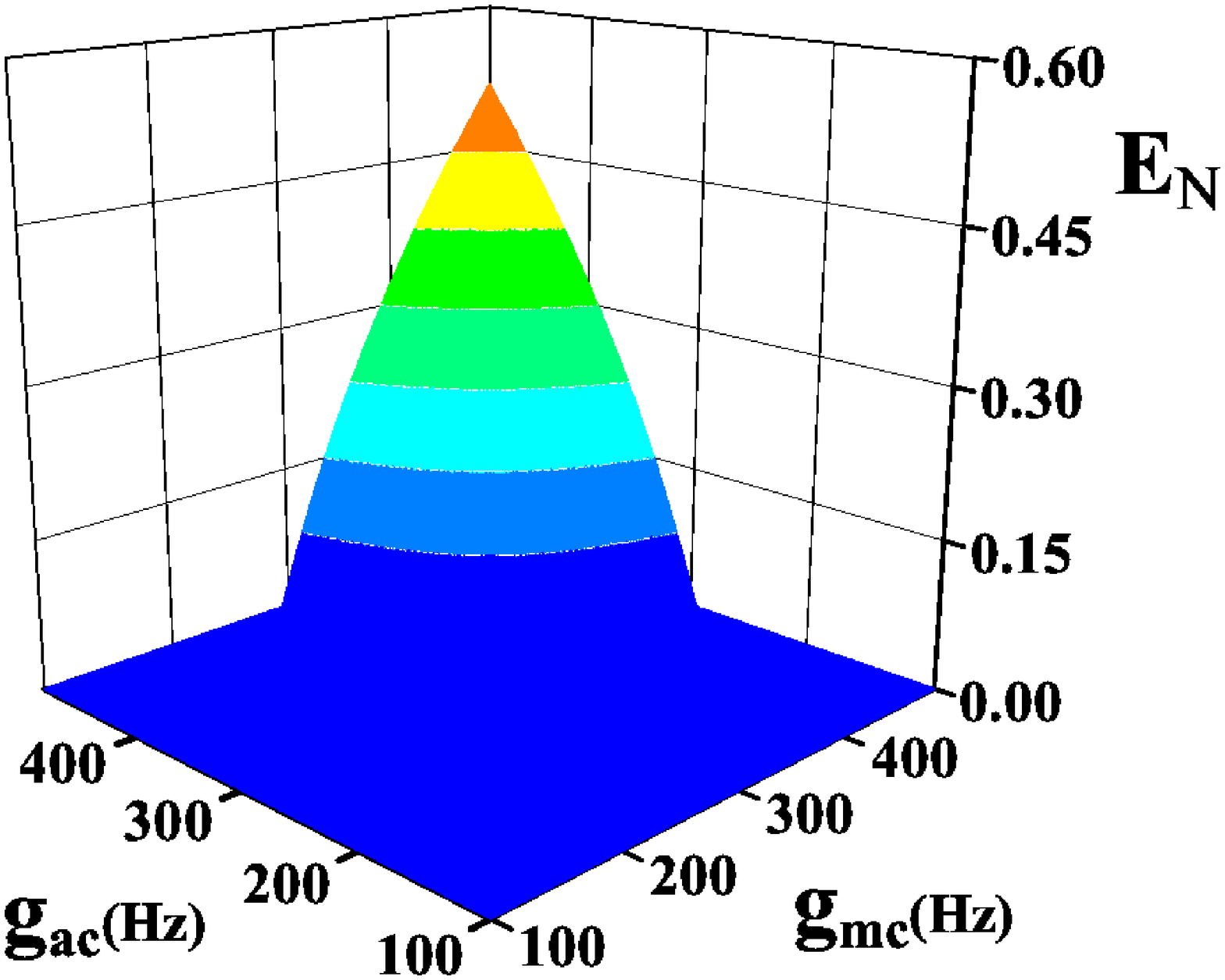}\label{e=2}}
 \subfigure[]{\includegraphics[scale=.158]{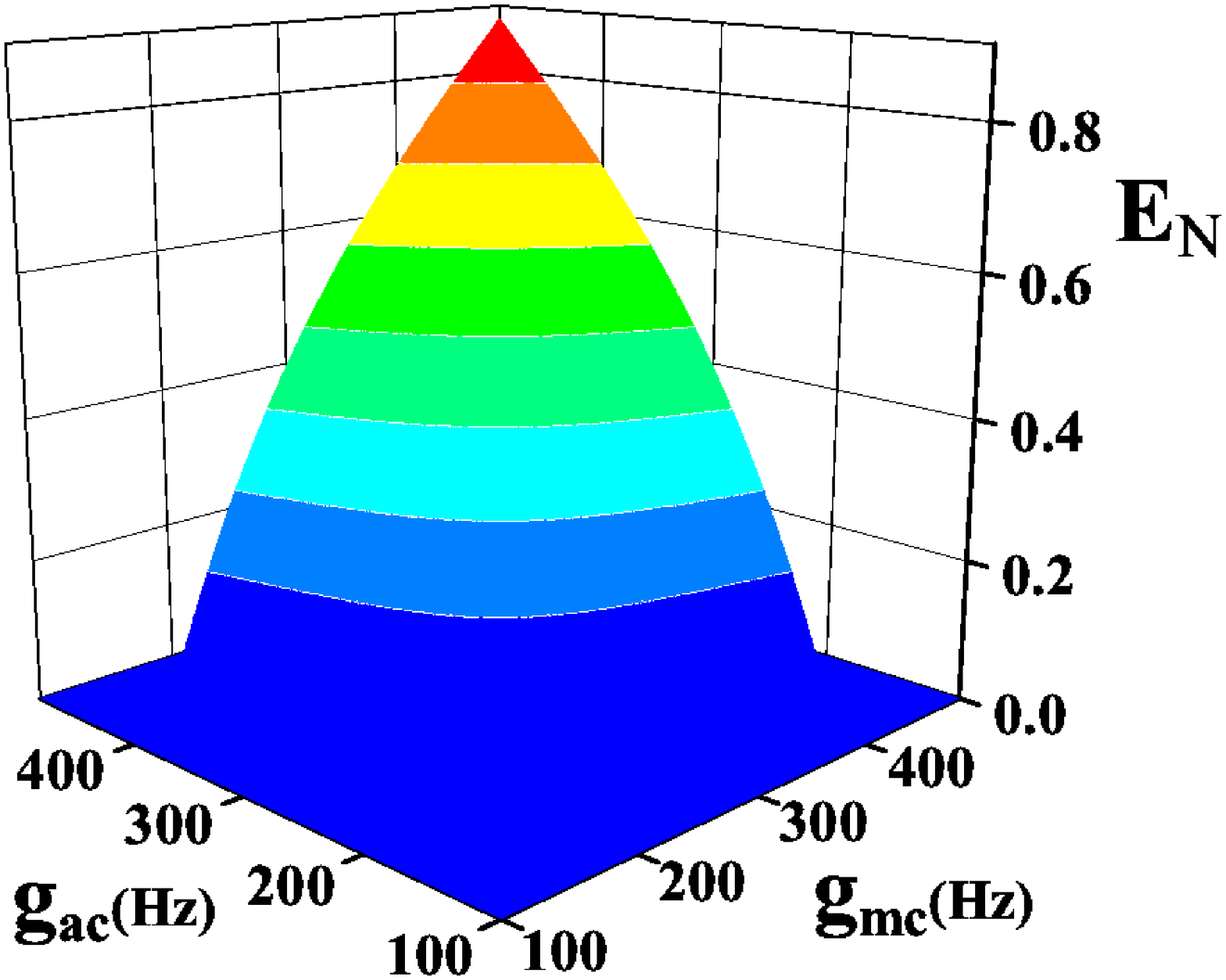}\label{e=3}}
\caption{ (Color online) (a) Plot of logarithmic negativity $\mathrm{E}_{\mathrm{N}}$ as a function of coupling of the intracavity mode with nano-mechanical resonator $\mathrm{g}_{\mathrm{mc}}$ and BEC mode $\mathrm{g}_{\mathrm{ac}}$ for $\Delta=2\pi\times 2\,\mathrm{MHz}$. The optical cavity length $\mathrm{L} = 1\, \mathrm{mm}$, driven by a laser with wavelength $\lambda =1000 \,\mathrm{nm}$, power $\mathrm{P} = 50\,\mathrm{mW}$ and $\omega_{\mathrm{m}} \simeq \omega_{\mathrm{a}}$. The mechanical mirror has a frequency $\omega_\mathrm{m}/2\pi=1\,\mathrm{MHz}$, and damping rate $\gamma=2\pi\times 100 \, \mathrm{Hz}$, its temperature is $\mathrm{T}=10\, \mathrm{\mu K}$ and cavity finesse is $\mathrm{F}=1\times10^4$. (b) $\mathrm{E}_{\mathrm{N}}$ versus $\mathrm{g}_{\mathrm{mc}}$  and  $\mathrm{g}_{\mathrm{ac}}$ for $\Delta=2\pi\times 3\,\mathrm{MHz}$.} \label{ent}
\end{figure}  
\paragraph*{}  Measurement of the entanglement between the mechanical resonator and Bose-Einstein condensate, requires as to compute $\mathrm{E}_\mathrm{N}$, which is obtained by tracing out the cavity mode, \textit{i.e}, removing the rows and columns of $\mathrm{V}$ which correspond to the cavity mode. The reduce state is still Gaussian and fully characterized by $4\times4$ matrix $ \mathrm{V'}$. In order to measure the entanglement between mechanical mirror and intracavity filed, we consider the Logarithmic negativity $\mathrm {E}_{\mathrm{N}}$. In the case of continuous variable (CV), $\mathrm{E}_{\mathrm{N}}$ can be defined as \cite{Adesso}
\begin{equation}
\mathrm{E}_{\mathrm{N}}=\mathrm{max}\,[0,-\ln 2\,\nu_-]\,,\label{55}
\end{equation}
 where,  $\nu_-= 2^{-1/2}\{\sum(\mathrm{V'})-[\sum(\mathrm{V'})^2-4\det \mathrm{V'}]^{1/2}\}^{1/2}$, with $\sum(\mathrm{V'})\equiv \det \mathrm{X} + \det\mathrm{Y}-2\det\mathrm{Z}$, is the smallest symplectic eigenvalue. However, the second eigenvalue  $\nu_+= 2^{-1/2}\{\sum(\mathrm{V'})+[\sum(\mathrm{V'})^2-4\det \mathrm{V'}]^{1/2}\}^{1/2}>>1/2$ at any value of the parameters. Therefore, it has no effect on the non-separability of the state \cite{Simon}. Moreover, the correlation matrix $\mathrm{V'}$ in $2\times2$ block form can be written as 
 \begin{center}
$\mathrm{V'}$=\(
 \begin{bmatrix}
\mathrm{X} & \mathrm{Z} \\ \mathrm{Z}^{\mathrm{t}} & \mathrm{Y}
 \end{bmatrix}
\).
\end{center}
It is clear from Eq.(\ref{55}) that the $\mathrm{E}_{\mathrm{N}}$ is the decreasing function of $,\nu_-$ and it quantifies how much two Gaussian states are entangled. The Gaussian state gets entangled only if $\nu_-<1/2$, and it is  Simon's necessary and sufficient entanglement non-positive partial transpose criterion of the Gaussian states \cite{Simon}, and this condition can also be written as $4\det \mathrm{V'}<\sum(\mathrm{V'})-1/4$.
\paragraph*{} In Fig.\,\ref{ent} we show that the entanglement between nano-mechanical resonator and Bose-Einstein condensate quantify by Logarithmic negativity $\mathrm{E}_{\mathrm{N}}$ as a function of the coupling of the mechanical resonator and Bose-Einstein condensate with field in the cavity. In Fig.\,\ref{e=2}, we have $\mathrm{E}_{\mathrm{N}}\simeq0.15$ for $\mathrm{g}_{\mathrm{mc}}=\mathrm{g}_{\mathrm{ac}}\thicksim 300\, \mathrm{Hz}$ and $\Delta= 2\pi\times2\,\mathrm{MHz}$. In Fig.\,\ref{e=3}, we find that the $\mathrm{E}_{\mathrm{N}}\simeq0.2$ for $\mathrm{g}_{\mathrm{mc}}=\mathrm{g}_{\mathrm{ac}}\thicksim 300\, \mathrm{Hz}$ and $\Delta= 2\pi\times3\,\mathrm{MHz}$. 
 \begin{figure}[ht]
\centering
\subfigure{\includegraphics[scale=.158]{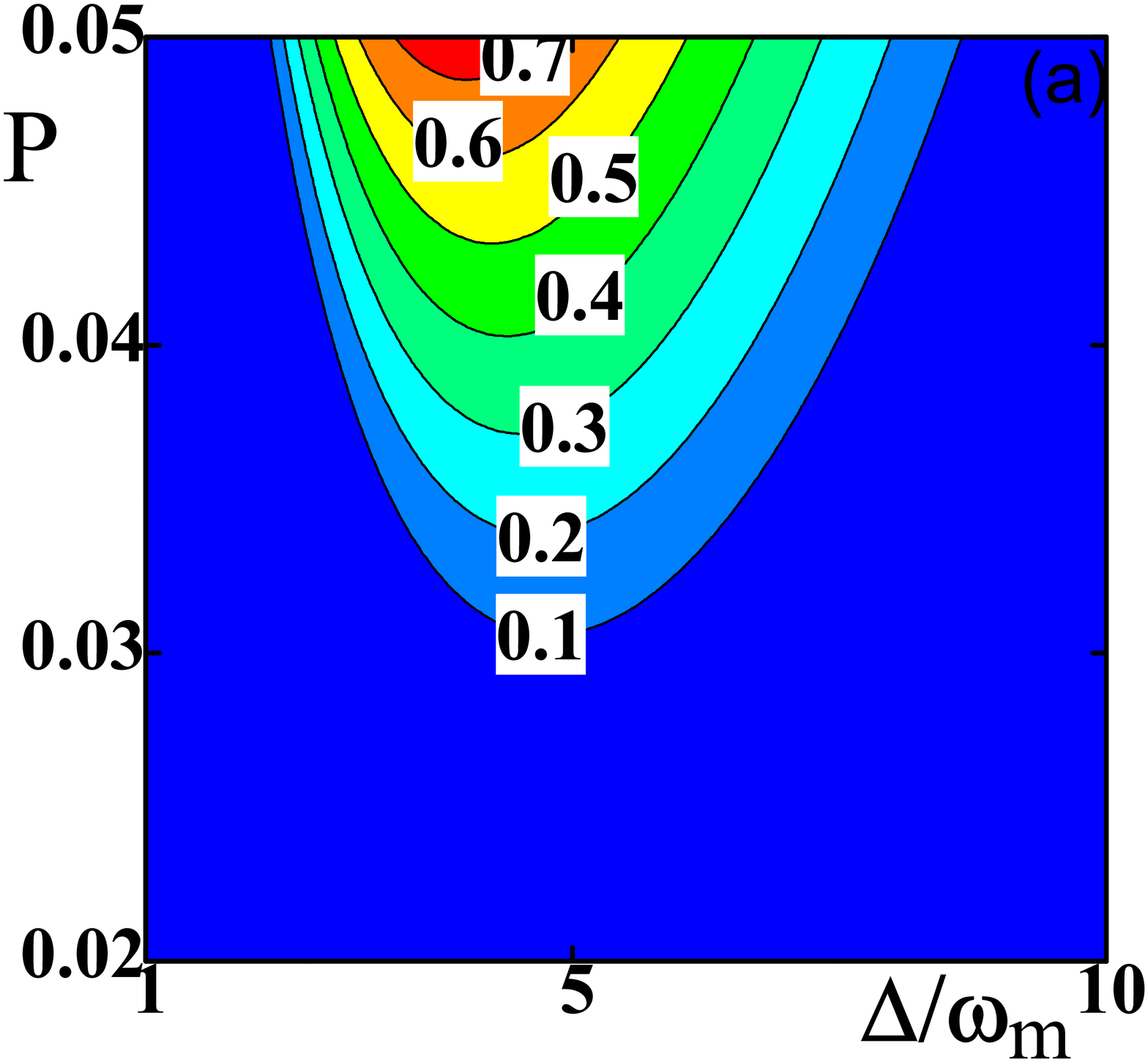}\label{P}}
\subfigure{\includegraphics[scale=.159]{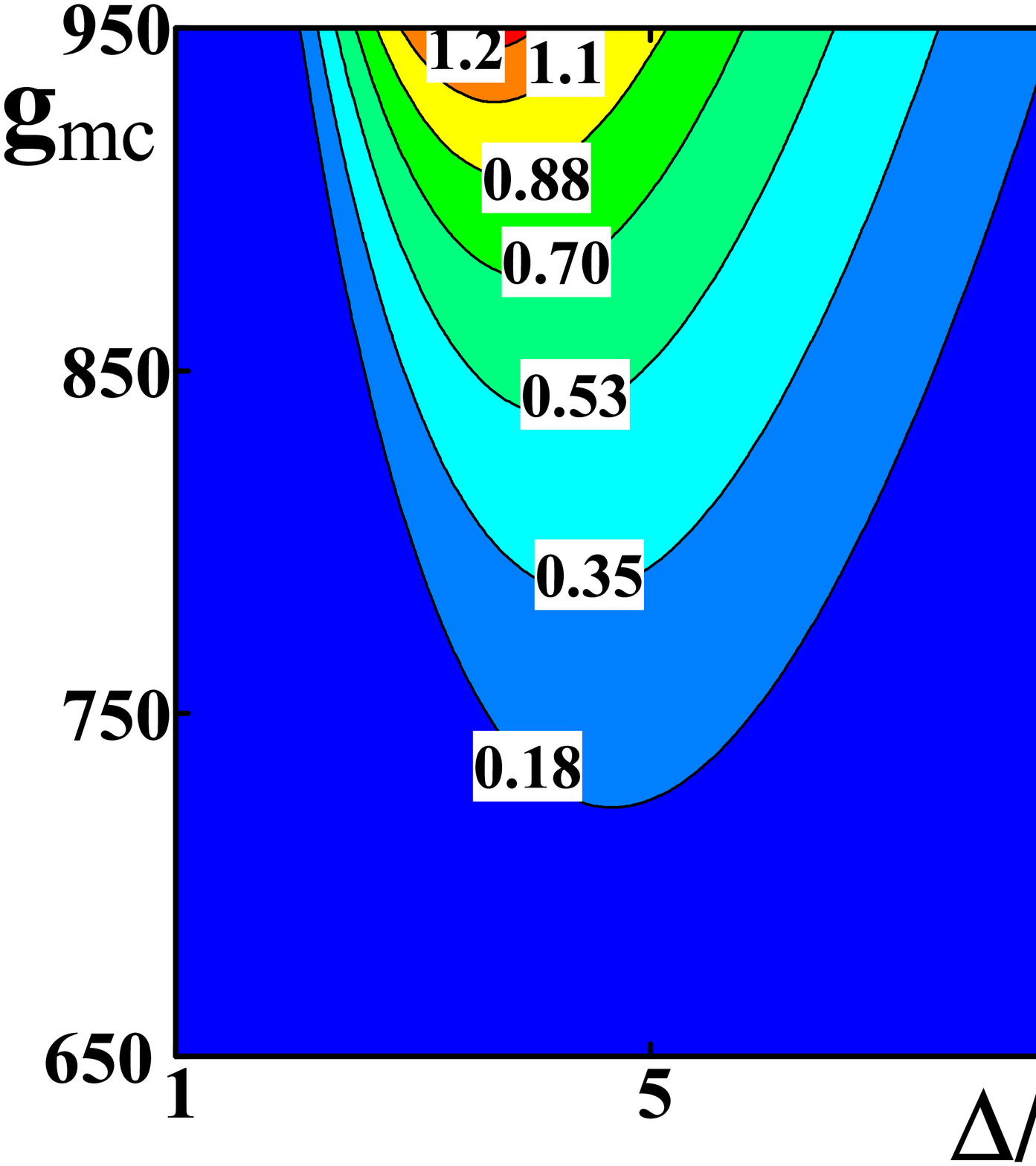}\label{z}}\\
\subfigure{\includegraphics[scale=.1565]{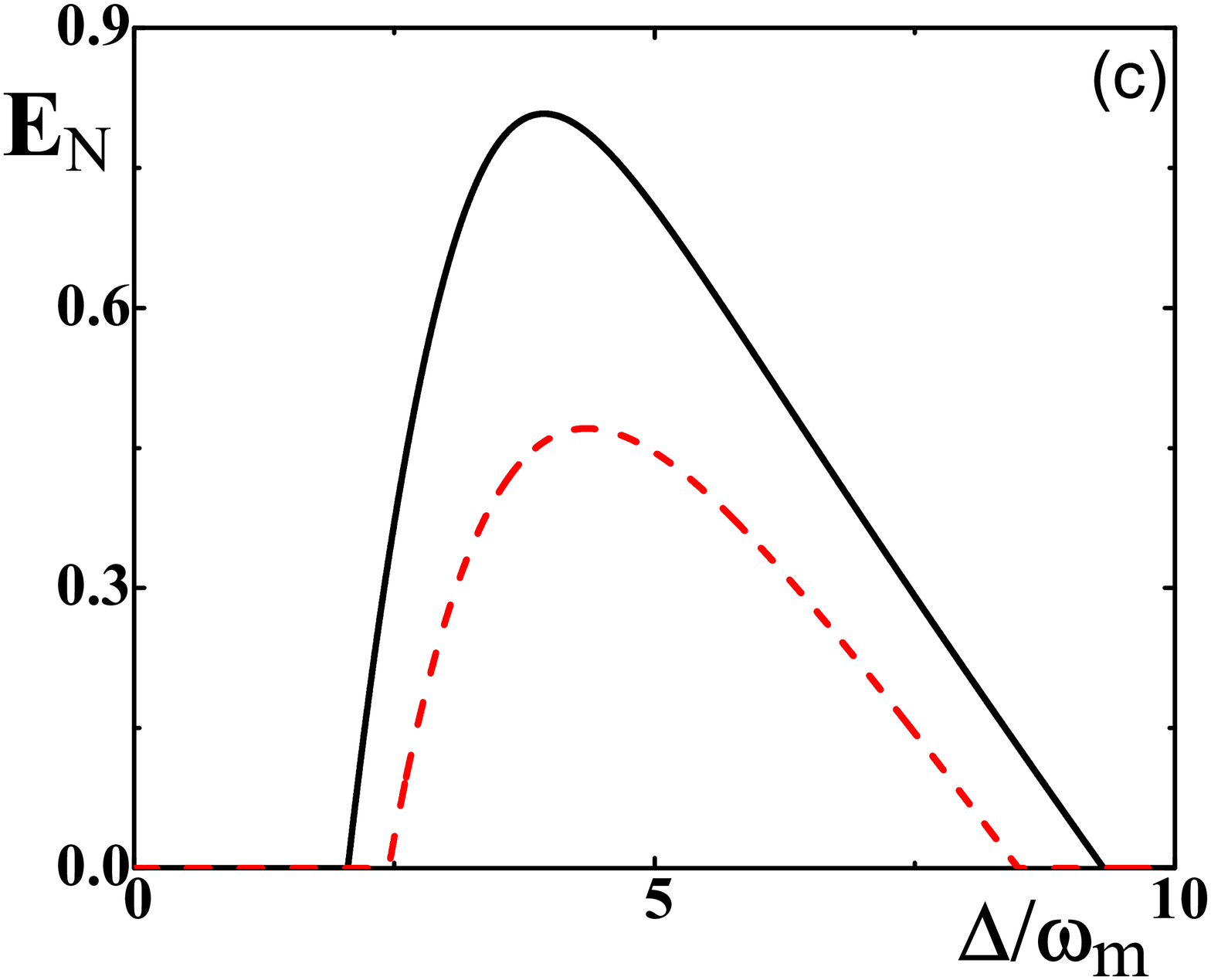}\label{e}}
\subfigure{\includegraphics[scale=.16]{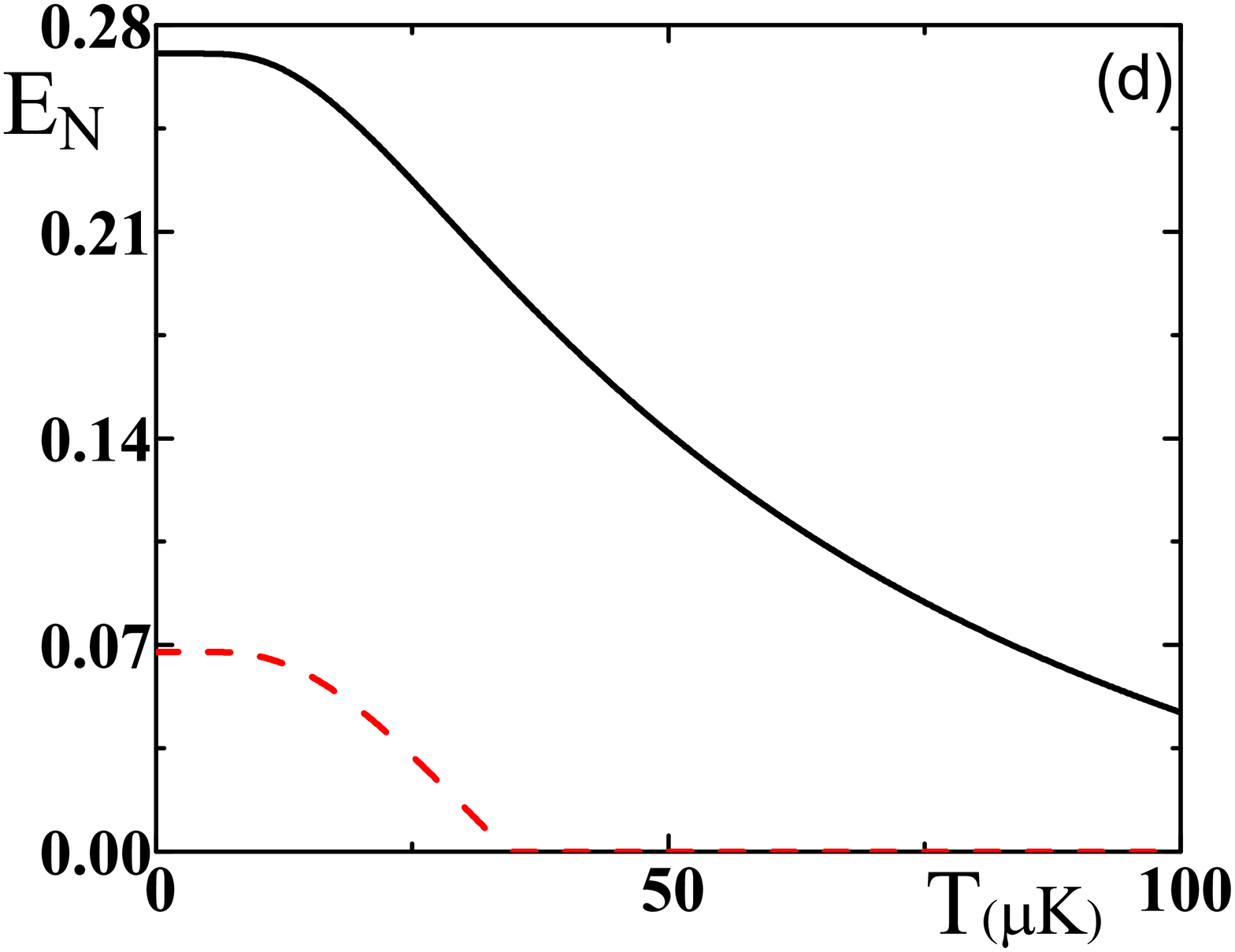}\label{T}}
\caption{(Color online) (a) Logarithmic negativity $\mathrm{E}_{\mathrm{N}}$ as a function of power of the input laser field and normalized detuning $\Delta/\omega_{\mathrm{m}}$. (b) $\mathrm{E}_{\mathrm{N}}$ as a function of coupling strength between mechanical mirror and intracavity field, \textit{i.e}, $\mathrm{g}_{\mathrm{mc}}$ and normalized detuning $\Delta/\omega_{\mathrm{m}}$ for a constant value of the coupling strength between BEC and intracavity field, \textit{i.e}, $\mathrm{g}_{\mathrm{ac}}=100\,\mathrm{Hz}$. (c) $\mathrm{E}_{\mathrm{N}}$ as a function of $\Delta/\omega_{\mathrm{m}}$ for  $\mathrm{m} = 4\,\mathrm{ng}$(black solid), $\mathrm{m} = 5\,\mathrm{ng}$(red dashed) and $\mathrm{g}_{\mathrm{ac}}=100\,\mathrm{Hz}$. (d) $\mathrm{E}_{\mathrm{N}}$ as a function of temperature T for $\Delta\simeq4\,\omega_{\mathrm{m}}$, $\mathrm{m} = 4\,\mathrm{ng}$ (black solid), $\mathrm{m} = 5\,\mathrm{ng}$ (red dashed) and $\mathrm{g}_{\mathrm{ac}}=100\,\mathrm{Hz}$. All the others parameters are the same as in Fig.\,\ref{ent}.} \label{pzet} 
\end{figure}
\paragraph*{} In Fig.\,\ref{pzet}, we also study the robustness of the steady state entanglement of the nano-mechanical resonator to the Bose-Einstein condensate with respect to several parameters in system. Fig.\,\ref{P} shows the dependence of $\mathrm{E}_{\mathrm{N}}$ on the input laser power $\mathrm{P}$ and normalized detuning $\Delta/\omega_{\mathrm{m}}$ for $\mathrm{g}_{\mathrm{ac}}=100\, \mathrm{Hz}$. Fig.\,\ref{z} shows the dependence of $\mathrm{E}_{\mathrm{N}}$ on optomechanical coupling rate $\mathrm{g}_{\mathrm{mc}}$ and normalized detuning $\Delta/\omega_{\mathrm{m}}$ for $\mathrm{g}_{\mathrm{ac}}=100\, \mathrm{Hz}$. Moreover, in Fig.\,\ref{e} we plot the  $\mathrm{E}_{\mathrm{N}}$ as function of normalized detuning $\Delta/\omega_{\mathrm{m}}$ for two different values of the mass m of the mechanical resonator, \textit{i.e}, black solid and red dashed lines for $\mathrm{m} = 4\,\mathrm{ng}$ and $\mathrm{m} = 5\,\mathrm{ng}$ respectively. One can easily observe that $\mathrm{E}_{\mathrm{N}}$ exists only within a finite interval of values of the $\Delta$ around $\Delta\simeq4.5\,\omega_{\mathrm m}$. Now, we investigate the effects of temperature on the entanglement. We consider $\mathrm{E}_{\mathrm{N}}$ of the evolved system at a fixed value of detuning. At high temperature, thermal fluctuation always suppress the entanglement of the coupling systems. Therefore, the logarithmic negativity is the decreasing function of temperature. In Fig.\,\ref{T}, we plot the logarithmic negativity as a function of temperature at a fixed value of the detuning $\Delta=2\pi\times4\,\mathrm{MHz}$. The black solid line refers to mass $\mathrm{m} = 4\,\mathrm{ng}$ and red dashed line refers to mass $\mathrm{m} = 5\,\mathrm{ng}$. Fig.\,\ref{T} shows that $\mathrm{E}_{\mathrm{N}}$ monotonically decreases with temperature. Moreover, it is noted that $\mathrm{E}_{\mathrm{N}}$ is also very sensitive to the mass of mechanical resonator and quickly decay for large value of the mass. 
\paragraph*{} Latter, we find the effective coupling between mechanical resonator and Bose-Einstein condensate. We consider the regime in which the dynamics of the cavity mode remains unperturbed due to the motion of the mechanical mode and BEC mode, and slowly mediates the interaction between the two. Therefore, we chose the regime where the cavity mode can be eliminated adiabatically. The Hamiltonian corresponding to the linearized quantum Langevin equations for the fluctuations operator $\partial\mathrm{q}_{\mathrm{m}},\, \partial\mathrm{p}_{\mathrm{m}},\,\partial\mathrm{q}_{\mathrm{a}},\,\partial\mathrm{p}_{\mathrm{a}},\partial\mathrm{q}_{\mathrm{c}}$ and $\partial\mathrm{p}_{\mathrm{c}}$ is given by 
\begin{eqnarray}
\mathcal{H}&=&\dfrac{\omega_{\mathrm m}}{2}\,\left(\partial\mathrm{p}^2_{\mathrm{m}} +\partial\mathrm{q}^2_{\mathrm{m}}\right) + \dfrac{\omega_{\mathrm a}}{2}\,\left(\partial\mathrm{p}^2_{\mathrm{a}} + \partial\mathrm{q}^2_{\mathrm{a}}\right)\nonumber \\
&& - \mathrm{G}_{\mathrm{mc}}\,\partial\mathrm{q}_{\mathrm{c}}\, \partial\mathrm{q}_{\mathrm{m}} + \mathrm{G}_{\mathrm{ac}}\,\partial\mathrm{q}_{\mathrm{c}}\, \partial\mathrm{q}_{\mathrm{a}}. \label{lh}
\end{eqnarray} 
 The corresponding parameteric regime for the fast cavity mode dynamics is, $\Delta>> \mathrm{G}_{\mathrm{mc}},\mathrm{G}_{\mathrm{ac}}$ or $\kappa >> \mathrm{G}_{\mathrm{mc}},\mathrm{G}_{\mathrm{ac}}$. Due to the second condition, cavity-mediated coherent dynamics is destroyed. Therefore, we only consider the regime where $\Delta$ takes large values, and the fluctuation quadratures of the cavity mode adiabatically follow the dynamics of positions fluctuations of the mechanical resonator and BEC mode. We further assume that both the mechanical resonator and BEC modes to be on resonance, \textit{i.e}, $\omega_{\mathrm a}=\omega_{\mathrm m}=\omega$. On eliminating the photon degree of freedom adiabatically we get the following effective Hamiltonian
 \begin{eqnarray}
\mathcal{H}_{\mathrm{eff}}&=& \mathcal{H}_{\mathrm{o}}+\mathcal{H}_{\mathrm{ma}} \nonumber \\
& =& \dfrac{\omega}{2} \,\left(\partial\mathrm{p}^2_{\mathrm{m}}+ \partial\mathrm{p}^2_{\mathrm{a}}\right) + \dfrac{\omega + \omega_{1}}{2}\, \partial\mathrm{q}^2_{\mathrm{m}} + \dfrac{\omega + \omega_{2}}{2}\, \partial\mathrm{q}^2_{\mathrm{a}} \nonumber\\
 &&+ \dfrac{\mathrm{G_{ma}}}{2} \,\partial\mathrm{q}_{\mathrm{m}}\partial\mathrm{q}_{\mathrm{a}}\,, \label{eff} 
\end{eqnarray}
where, $\omega_{\mathrm{1}} = 4\,\mathrm{G}^2_{\mathrm{mc}}\,\Delta/(\kappa^2 + \Delta^2)$, $\omega_{\mathrm{2}} = 4\,\mathrm{G}^2_{\mathrm{ac}}\,\Delta/(\kappa^2 + \Delta^2)$ and $\mathrm{G_{ma}} = -8\,\mathrm{G}_{\mathrm{ac}}\,\mathrm{G}_{\mathrm{mc}}\,\Delta/(\kappa^2+\Delta^2)$. It is noted that the effective interaction $\mathrm{G}_{\mathrm{ma}}$ between mechanical resonator and Bose-Einstein condensate is increased as the coupling of mechanical oscillator and BEC is increased with the intracavity field.    
The effective interaction between a mechanical resonator and Bose-Einstein condensate can be described via the Hamiltonian $\mathcal{H}_{\mathrm{ma}}$, from  Eq.(\ref{eff}) 
\begin{equation}
\mathcal{H}_{\mathrm{ma}}=\mathrm{G_{ma}}(\partial \mathrm{a}\,\partial \mathrm{b}^\dag + \partial \mathrm{a}^\dag\,\partial \mathrm{b})/2\,+\, \mathrm{G_{ma}}(\partial \mathrm{a}\,\partial\mathrm{b}\, +\, \partial \mathrm{a}^\dag\,\partial \mathrm{b}^\dag)/2, \label{squeez} 
\end{equation}
 where $\mathrm{G_{ma}}$ is the effective coupling strength between mechanical mode and BEC. This Hamiltonian is analogous to the interaction of two optical fields in a nonlinear medium generating the 2-modes squeezed state \cite{Ou}. The first term in Eq.(\ref{squeez}) corresponds to an energy exchange take place between mechanical and atomic modes. The second term in Eq.(\ref{squeez}) accounts for the down-conversion interaction, describes the creation and annihilation of the atomic and mechanical modes phonons in pairs, which corresponds to entangling the atomic and mechanical modes.  
For $\mathrm{G_{ma}}\simeq\omega_{\mathrm{m}}$ or greater, the mechanical resonator and Bose-Einstein condensate are entangled, and the parameters are chosen such that this condition is fulfilled.      
\section{\bf{Experimental detection}}
 For experimental realization of the generated entanglement, one has to measure several quadrature correlations \cite{Duan} as have been experimentally measured for the entanglement of the two optical modes \cite{Laurat}. However, in our case we consider another Fabry Perot cavity adjacent to first one and is driven by a weak laser field, a scheme of this kind has been discussed in  \cite{Vitali}. In addition, we assume that the movable mirror is perfect reflector at both sides so there is entanglement between the optical modes of the two cavities. The equation of motion of the annihilation operator $\mathrm{c_1}$ of the optical mode of the second cavity similar to the linearized version of Eq.(\ref{linrz}) is  
\begin{eqnarray}
\partial\dot{\mathrm{c}}_1 &=& -\left(\kappa_1+i\,\Delta_1 \right)\partial \mathrm{c}_1 +i\dfrac{\mathrm{G}_{\mathrm{mc1}}}{2}\left(\partial\mathrm{b}+\partial\mathrm{b}^\dag\right)\nonumber\\
&&-i\dfrac{\mathrm{G}_{\mathrm{ac1}}}{2}\left(\partial\mathrm{a}\,+\,\partial\mathrm{a}^\dag\right)+\sqrt{2\,\kappa_1}\, \mathrm{c}_{\mathrm{in1}}\,, \label{c2}
\end{eqnarray}
 where $\kappa_1$, $ \Delta_1 $ and $\mathrm{c_{in_1}}$ are the cavity decay rate, effective detuning and the input noise of the second cavity mode respectively. In addition, $\mathrm{G_{mc1}}$ and  $\mathrm{G_{ac1}}$ are the effective coupling rates of the second cavity mode to the mechanical resonator and Bose-Einstein condensate respectively. Moreover, we assume $\left<\mathrm{c}\right>_\mathrm{ss} >> \left<\mathrm{c}\right>_\mathrm{ss1}$ and $\Delta_1=\omega_{\mathrm{m}}>>\kappa$, $\mathrm{G_{mc1}}$, $\mathrm{G_{ac1}}$. Therefore, in the rotating frame at $\Delta_1=\omega_{\mathrm m}$, Eq.(\ref{c2}) for slow variables $\partial \tilde{o}(t)=\partial o(t)\mathrm{exp}\left(i\omega_{\mathrm{m}}t\right) $ can be rewritten as
\begin{eqnarray}
\partial\dot{\tilde{\mathrm{c}}}_1 &=& -\kappa_1 \partial \tilde{\mathrm{c}}_1 + \dfrac{i}{2}\left[\mathrm{G_{mc1}}\partial\tilde{\mathrm{b}}-\mathrm{G_{ac1}}\partial\tilde{\mathrm{a}}\right]\nonumber\\
&&+\sqrt{2\,\kappa_1} \tilde{\mathrm{c}}_{\mathrm{in1}}. \label{slow}
\end{eqnarray}
Here, we assume $\omega_{\mathrm{m}}=\omega_{\mathrm{a}}=\omega$ and neglect the fast oscillating terms at frequency $2\omega$. On eliminating the cavity mode adiabatically, Eq.(\ref{slow}) can be written as
\begin{equation}
\partial\tilde{\mathrm{c}}_1\simeq \dfrac{i}{2\kappa_1}\left[\mathrm{G_{mc1}}\partial\tilde{b} - \mathrm{G_{ac1}}\partial\tilde{a}\right]+ \sqrt{\dfrac{2}{\kappa_1}}\tilde{\mathrm{c}}_{\mathrm{in1}}.                     
\end{equation}
According to the stranded input-output relation \cite{Gardiner}, $\tilde{\mathrm c}_{\mathrm{out1}}=\sqrt{2\,\kappa_1}\,\partial\tilde{\mathrm {c}}_1- \tilde{\mathrm{c}}_{\mathrm{in1}}$, the output field is given by
\begin{equation}
\tilde{\mathrm c}_{\mathrm{out1}}=\dfrac{i}{2\kappa_1}\left[\mathrm{G_{mc1}}\partial\tilde{b} - \mathrm{G_{ac1}}\partial\tilde{a}\right] + \tilde{\mathrm{c}}_{\mathrm{in1}}. \label{homo}
\end{equation}
From Eq.(\ref{homo}) we can measure the $\mathrm{G_{mc1}}\partial\tilde{b} - \mathrm{G_{ac1}}\partial\tilde{a}$ by homodyning the output of the second cavity mode. Therefore, all the entries of the correlation matrix can be determined by measuring the correlation between the output of the two cavities, hence from the correlation matrix logarithmic negativity can be calculated numerically.  
\section{\bf{Conclusion}}      
In conclusion, we discuss quantum correlation in a system which consists of Bose-Einstein condensates trapped inside a Fabry-Perot cavity with a moving end mirror driven by single mode optical field. We describe a scheme to generate the steady state entanglement of the motional degree of freedom of the nano-mechanical resonator (end mirror of the cavity) and collective oscillation of the atomic density of the Bose-Einstein condensate. Moreover, it is observed that the entanglement generated between Bose-Einstein condensate and nano-mechanical resonator is very sensitive with respect to temperature and persist only upto $20\,\mathrm{\mu K}$ and $100 \,\mathrm{\mu K}$ for mass $4\, \mathrm{ng}$ and $5\, \mathrm{ng}$, respectively.  The generated entanglement is measured by considering the second optical cavity driven by a weak laser field.          

\end{document}